\documentclass[11pt,preprint]{aastex}

\shorttitle{Fundamental Parameters of $\epsilon$ Eridani}
\shortauthors{Baines \& Armstrong}

\usepackage{amsmath}
\begin{document}

%\journalinfo{Astrophysical Journal, accepted}
%\submitted{Astrophysical Journal, accepted}

\title{CONFIRMING FUNDAMENTAL PROPERTIES OF THE EXOPLANET HOST STAR $\epsilon$ ERIDANI USING THE NAVY OPTICAL INTERFEROMETER}

\author{Ellyn K. Baines, J. Thomas Armstrong} % Henrique?
\affil{Remote Sensing Division, Naval Research Laboratory, 4555 Overlook Avenue SW, \\ Washington, DC 20375}
\email{ellyn.baines@nrl.navy.mil, tarmstr@crater.nrl.navy.mil}

\begin{abstract}

We measured the angular diameter of the exoplanet host star $\epsilon$ Eridani using the Navy Optical Interferometer. We determined its physical radius, effective temperature, and mass by combining our measurement with the star's parallax, photometry from the literature, and the Yonsei--Yale isochrones \citep{2001ApJS..136..417Y}, respectively. We used the resulting stellar mass of 0.82$\pm$0.05 $M_\odot$ plus the mass function from \citet{2006AJ....132.2206B} to calculate the planet's mass, which is 1.53$\pm$0.22 $M_{\rm Jupiter}$. Using our new effective temperature, we also estimated the extent of the habitable zone for the system.

\end{abstract}


%%%%%%%%%%%%%%%%%%%%%%%%%%% Introduction %%%%%%%%%%%%%%%%%%%%%%%%%%
\section{Introduction}

\citet{2000ApJ...544L.145H} discovered a planet orbiting $\epsilon$ Eri (HD 22049, K2 V) using radial velocity measurements that spanned 20 years. They determined the planet's orbit and a minimum mass of 0.86 $M_{\rm Jupiter}$. They assumed a stellar mass of 0.85 $M_\odot$ from \citet{1993ApJ...412..797D}, who used spectroscopic observations to derive parameters such as effective temperature, surface gravity, metallicity, and luminosity and then used those parameters to calculate the star's mass.

\citet{2004AandA...426..601D} observed $\epsilon$ Eri using the Very Large Telescope Interferometer (VLTI) during the commissioning period of the VLT Interferometer Commissioning Instrument (VINCI). They measured a limb--darkened angular diameter of 2.148$\pm$0.010$\pm$0.027 mas with a global uncertainty of 0.029 mas, the last of which leads to an error of 1.4$\%$. The first two errors on the diameter are the statistical and systematic errors, respectively. The statistical error is based on the dispersion of the data while the systematic error is based on uncertainties in the calibrators' diameters. Di Folco et al. then used the CESAM code version 4 evolutionary models by \citet{1997AandAS..124..597M} to obtain various parameters for the host star, such as luminosity, effective temperature, age, mass, radius, surface gravity, and metallicity. The resulting CESAM mass for $\epsilon$ Eri was 0.83 $M_\odot$ while they list their estimated mass as 0.90$\pm$0.10 $M_\odot$ and the adopted mass as 0.84 $M_\odot$.

\citet{2006AJ....132.2206B} combined Hubble Space Telescope Fine Guidance Sensor observations of $\epsilon$ Eri with ground--based astrometric and radial velocity data to produce an orbit for the system and the mass of the planetary companion. They assumed a mass for the central star of 0.83 $M_\odot$ and a young age for the star, $\sim$800 Myr. Their resulting $M_{\rm planet}$ was 1.55$\pm$0.24 $M_{\rm Jupiter}$.

We used the Navy Optical Interferometer\footnote{The NOI underwent a recent name change. It was previously known as the Navy Prototype Optical Interferometer.} (NOI) to measure the angular diameter of $\epsilon$ Eri and then calculated its physical radius and effective temperature using the combination of our new measurement and other observed quantities, such as parallax, interstellar absorption, and bolometric correction. We provide an independent estimate of the star's mass by placing it on an H--R diagram using our values, which then leads to an estimate of the planet's mass. Section 2 describes the instrument and our observing procedure, Section 3 discusses how $\epsilon$ Eri's angular diameter, physical radius, and $T_{\rm eff}$ were determined, and Section 4 explores the physical implications of the new measurements including the mass determination and habitable zone calculation.

%%%%%%%%%%%%%%%%%%%%%%%%% Interferometric observations %%%%%%%%%%%%%%%%%%%%
\section{Interferometric observations}
Observations were obtained using the NOI, an interferometer located on Anderson Mesa, AZ \citep{1998ApJ...496..550A}. The NOI consists of two nested arrays: the four stations of the astrometric array (AC, AE, AW, and AN, which stand for astrometric center, east, west, and north, respectively) and the six stations of the imaging array, of which two stations (E6 and W7) are currently in operation. The baselines\footnote{The ``baseline'' is the distance between the stations.} range from 16 to 79 m in length and the siderostats are 50 cm in diameter, of which the central 12 cm are used. The NOI observes in 16 spectral channels simultaneously over the 550 to 850 $\mu$m range.

Each observation consisted of a 30--second coherent (on the fringe) scan in which the fringe contrast was measured every 2 ms, paired with an incoherent (off the fringe) scan used to estimate the additive bias affecting the visibility measurements \citep{2003AJ....125.2630H}. Scans were taken on five baselines simultaneously. Each coherent scan was averaged to 1--second data points, and then to a single 30--second average. The dispersion of 1--second points provided an estimate of the internal errors.

$\epsilon$ Eri was observed over three nights in 2010 using baseline lengths from 19 to 79 m (see Table \ref{calib_visy}). We interleaved observations of $\epsilon$ Eri with the calibrator star 38 Eri (HD 26574) so that every target was flanked by calibrator observations made as close in time as possible, which allowed us to convert instrumental target and calibrator visibilities to calibrated visibilities for the target. 38 Eri was selected because it appears to be a single star that is significantly more unresolved on the baselines used than $\epsilon$ Eri. This meant that uncertainties in the calibrator's diameter did not affect the target's diameter calculation as much as if the calibrator star had a substantial angular size on the sky.

To establish a diameter estimate for 38 Eri and to check for excess emission from unseen close companions that would make it unsuitable as a calibrator, we fit a spectral energy distribution (SED) to published $UBVRIJHK$ photometry, after converting magnitudes to fluxes using \citet{1996AJ....112..307C} for $UBVRI$ values and \citet{coh03} for $JHK$ values. We used Kurucz model atmospheres\footnote{Available to download at http://kurucz.cfa.harvard.edu.} with effective temperature ($T_{\rm eff}$) and surface gravity (log~$g$) values from the literature to calculate the SED and determine the angular diameter that best fit the photometry. Figure \ref{cal_sed} shows the SED fit. The photometry, $T_{\rm eff}$ and log~$g$ values, and the resulting diameter are listed in Table \ref{sed_params}.

%%%%%%%%%%%%%%%%% Angular diameter determinations %%%%%%%%%%%%%%%%
\section{Determination of angular diameter and $T_{\rm eff}$}
The observed quantity of an interferometer is defined as the visibility squared ($V^2$), which we fit with a model of a uniformly--illuminated disk (UD) that represents the face of the star. Diameter fits to $V^2$ were based upon the UD approximation given by $V^2 = [2 J_1(x) / x]^2$, where $J_1$ is the first--order Bessel function and $x = \pi B \theta_{\rm UD} \lambda^{-1}$, where $B$ is the projected baseline at the star's position, $\theta_{\rm UD}$ is the apparent UD angular diameter of the star, and $\lambda$ is the effective wavelength of the observation \citep{shao92}. A more realistic model of a star's disk involves limb--darkening (LD), and the relationship incorporating the linear LD coefficient $\mu_{\lambda}$ \citep{han74} is:
\begin{equation}
V^2 =  \left( {1-\mu_\lambda \over 2} + {\mu_\lambda \over 3} \right)^{-2}
\times
\left[(1-\mu_\lambda) {J_1(\rm x) \over \rm x} + \mu_\lambda {\left( \frac{\pi}{2} \right)^{1/2} \frac{J_{3/2}(\rm x)}{\rm x^{3/2}}} \right] ^2 .
\end{equation}
Table~\ref{calib_visy} lists the date of observation, spatial frequency (a function of wavelength and baseline), calibrated visibilities ($V^2$), and errors in $V^2$ ($\sigma V^2$) for $\epsilon$ Eri.\footnote{Table \ref{calib_visy} shows the information for a portion of one night as an example, and the full table is available on the electronic version of \emph{The Astrophysical Journal}.} 

\begin{figure}[h]
\includegraphics[width=0.75\textwidth, angle=90]{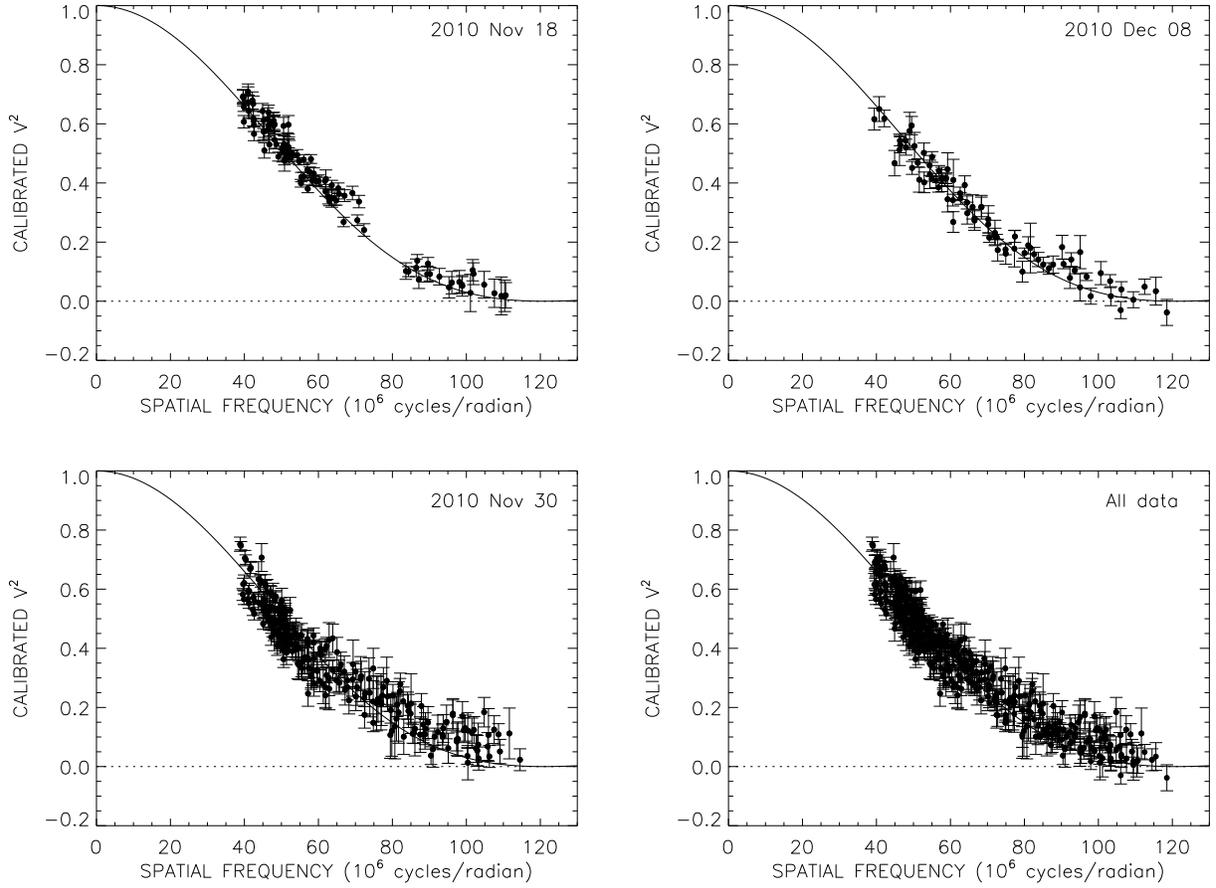}
\caption{$\epsilon$ Eri LD diameter fit for each night and all the data combined. The solid lines represent the theoretical visibility curves for the star with the best fit $\theta_{\rm LD}$, the filled circles are the calibrated visibilities, and the vertical lines are the measured errors.}
  \label{vvsb}
\end{figure}

The LD coefficient was obtained from \citet{cla95} after adopting the $T_{\rm eff}$ and log~$g$ values required for the star. The $T_{\rm eff}$ used to determine $\mu_{\lambda}$ has little effect on the final $\theta_{\rm LD}$: if $T_{\rm eff}$ varies by 500 K in either direction, the resulting change in $\theta_{\rm LD}$ of less than 1$\%$. Similarly, the final angular diameter is little affected by the choice of $\mu_{\lambda}$. A 10$\%$ change in the $\mu_{\lambda}$ leads to a change in the measured LD diameter of less than $1\%$. Figure \ref{vvsb} shows the LD diameter ($\theta_{\rm LD}$) fit for $\epsilon$ Eri for each night as well as the fit to all the data simultaneously. The last results in an angular diameter of 2.153$\pm$0.028 mas\footnote{The fit for each individual night was within 3$\%$ of the LD diameter fit that incorporated all the data.}. We then combined our measurement of the star's $\theta_{\rm LD}$ and the parallax determined by \citet{2006AJ....132.2206B} to calculate the star's physical radius: 0.74$\pm$0.01 $R_\odot$. Table \ref{parameters} lists the measurements made here as well as other stellar parameters.

\begin{figure}[h]
\includegraphics[width=1.0\textwidth]{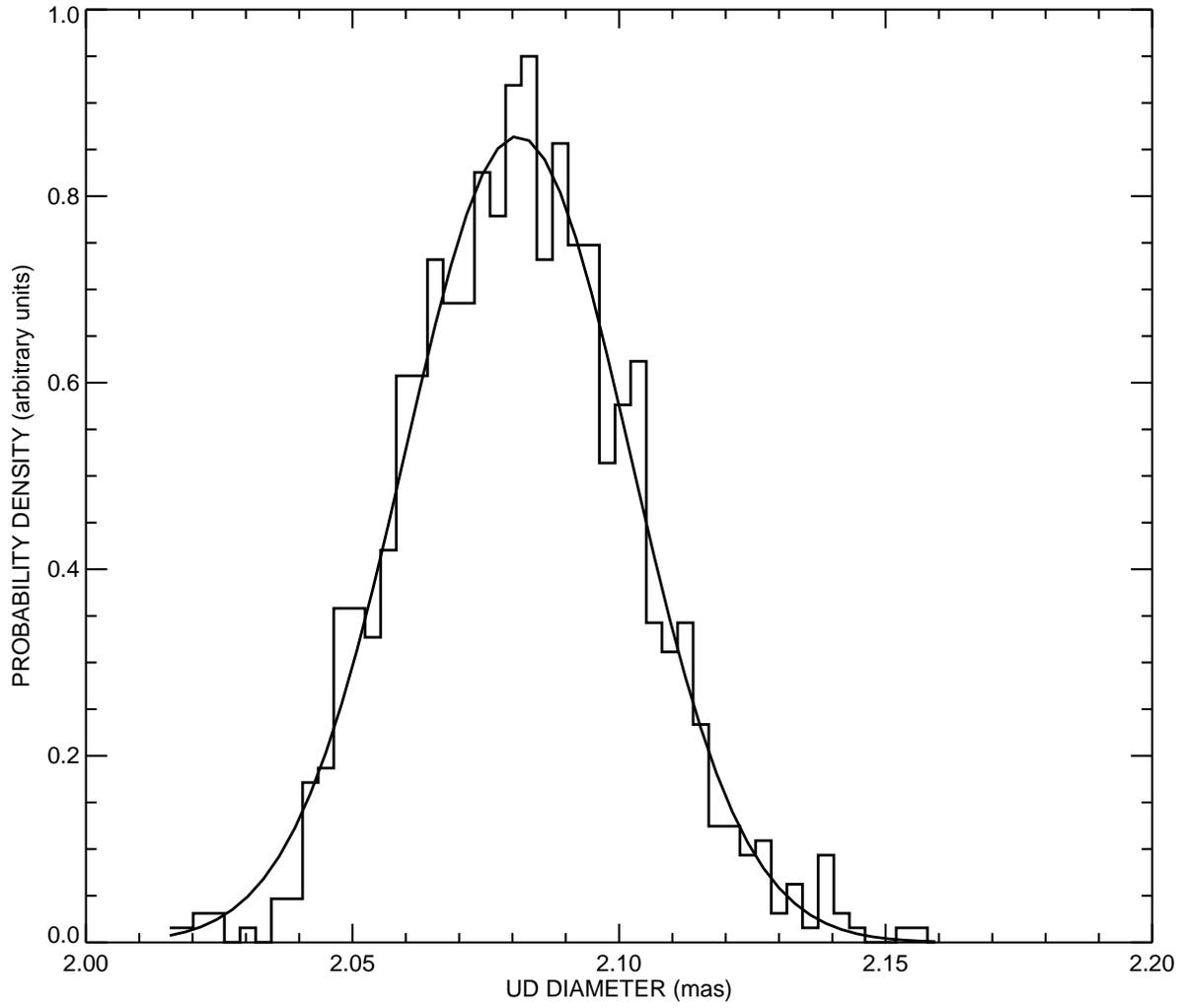}
\caption{Histogram representing the probability density function for $\epsilon$ Eri's LD diameter fit. A Gaussian fit to the distribution is also shown as the solid curve.}
  \label{error_dist}
\end{figure}

The error for the LD diameter fit was derived using the method described in \citet{2010SPIE.7734E.103T}, who showed that a non-linear least-squares method does not sufficiently account for atmospheric effects on time scales shorter than the window between target and calibrator observations. They describe a bootstrap Monte Carlo method that treats the observations as groups of data points because the NOI collects data in scans consisting of 16 channels simultaneously. They discovered that when the data points were analyzed individually, a single scan's deviation from the trend had a large impact on the resulting diameter and error calculation. On the other hand, when they preserved the inherent structure of the observational data by using scans of 16 channels instead of individual data points, the uncertainty on the angular diameter was larger and more realistic. This method makes no assumptions about underlying errors due to atmospheric effects, which are applicable to all stars observed using ground--based instruments. Figure \ref{error_dist} shows the probability density function for $\epsilon$ Eri's LD diameter measurement.

Once $\theta_{\rm LD}$ was determined interferometrically, the $T_{\rm eff}$ was calculated using the relation 
\begin{equation}
F_{\rm BOL} = {1 \over 4} \theta_{\rm LD}^2 \sigma T_{\rm eff}^4,
\end{equation}
where $F_{\rm BOL}$ is the bolometric flux and $\sigma$ is the Stefan--Boltzmann constant. $F_{\rm BOL}$ was computed in the following way: the star's $V$ and $K$ magnitudes were dereddened using the extinction curve described in \citet{1989ApJ...345..245C} and the interstellar absorption ($A_{\rm V}$) value. The intrinsic broadband color ($V-K$) was calculated and the bolometric correction (BC) was determined by interpolating between the [Fe/H] = 0.0, and -1.0 tables from \citet{1999AandAS..140..261A}. They point out that in the range of 6000 K $\geq T_{\rm eff} \geq$ 4000 K, their BC calibration is symmetrically distributed around a $\pm$0.10 mag band when compared to other calibrations, so we assigned the BC an error of 0.10. $F_{\rm BOL}$ was determined by applying the BC and the $T_{\rm eff}$ was calculated to be 5039$\pm$126 K. The star's luminosity ($L$) was also calculated using the absolute $V$ magnitude and the BC. See Table \ref{parameters} for a list of all these parameters.

Because the BC is an important parameter in the $T_{\rm eff}$ determination, we also derived the BC using the equation relating log($T_{\rm eff}$) and BC presented by \citet{1996ApJ...469..355F} and updated by \citet{2010AJ....140.1158T}. We used a range of $T_{\rm eff}$ values that bracketed the $T_{\rm eff}$ listed for $\epsilon$ Eri in \citet{2004AJ....127.1227C} by 450 K. The maximum difference in BC between the Flower and Alonso et al. calculations was 0.03, which is well within our assigned error bar of 0.10, and only changed the final $T_{\rm eff}$ by a maximum 34 K (0.7$\%$).

%%%%%%%%%%%%%%%%%%%%%%%%%%% Results %%%%%%%%%%%%%%%%%%%%%%%%%%
\section{Results and discussion}
\begin{figure}[h]
\includegraphics[width=1.0\textwidth]{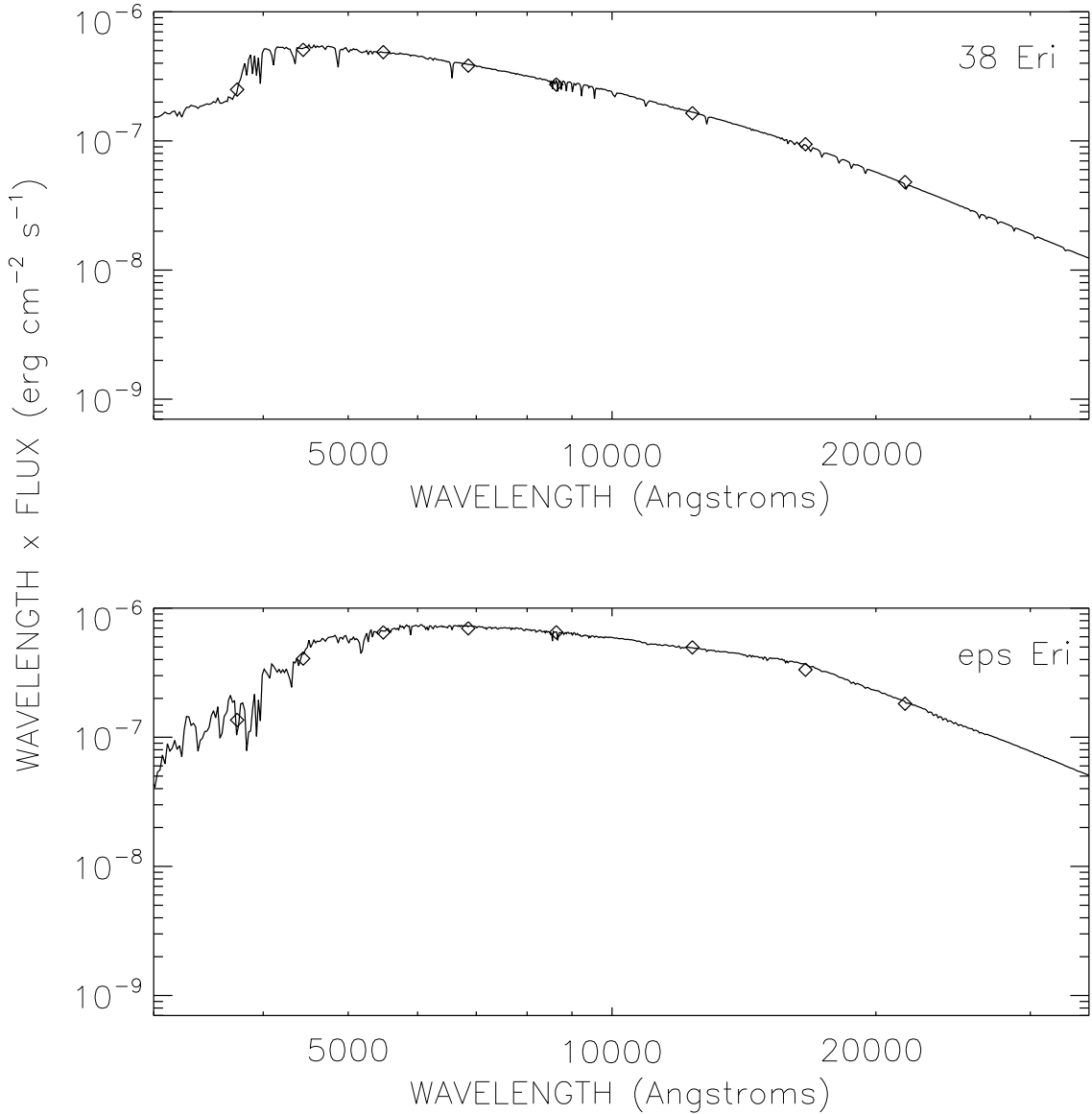}
\caption{SED fits for the calibrator star 38 Eri (top panel) and $\epsilon$ Eri (bottom panel). The diamonds are fluxes derived from $UBVRI JHK$ photometry (left to right) and the solid lines are the Kurucz stellar models of the stars. The errors for the $UBV$ measurements were $<1\%$, no errors were listed for $RI$, and the errors for $JHK$ were $7-16\%$, which are not indicated on the plot. See Table \ref{sed_params} for the values used to create the fits.}
  \label{cal_sed}
\end{figure}

As a check for our result we estimated the LD diameter for $\epsilon$ Eri using two additional methods. First, we created an SED fit for the star as described in Section 2 and Figure \ref{cal_sed} shows the result. The only free parameter for the SED fit is the angular diameter. Second, we used the relationship described in \citet{2004AandA...426..297K} between the ($V-K$) color and log $\theta_{\rm LD}$. Our measured $\theta_{\rm LD}$ is 2.153$\pm$0.028 mas, the SED fit estimates 2.04$\pm$0.11 mas, and the color--diameter relationship produces 2.04$\pm$0.82 mas.

The main sources of errors for the three methods are uncertainties in visibilities for the interferometric measurement, uncertainties in the comparison between observed fluxes and the model fluxes for a given $T_{\rm eff}$ and log~$g$ for the SED estimate, and uncertainties in the $K$ magnitude and in the parameters of the relation and the spread of stars around that relation for the color--diameter determination. All three diameters agree within their errors but our interferometric measurement provides an error 4 and 29 times smaller than the other methods, respectively. 

Our measurement is consistent with that derived using the FLUOR beam combining instrument on the Center for High Angular Resolution Astronomy Array at 2.126$\pm$0.014 mas \citep{2007AandA...475..243D}. FLUOR observes in the $K'$--band at $1.94 - 2.34$ $\mu$m and our measurement in visible wavelengths confirms their value. This was expected because both analyses incorporate limb darkening in their respective wavelengths, which will compensate for any difference due to the wavelength of the observations.

\begin{figure}[h]
\includegraphics[width=1.0\textwidth]{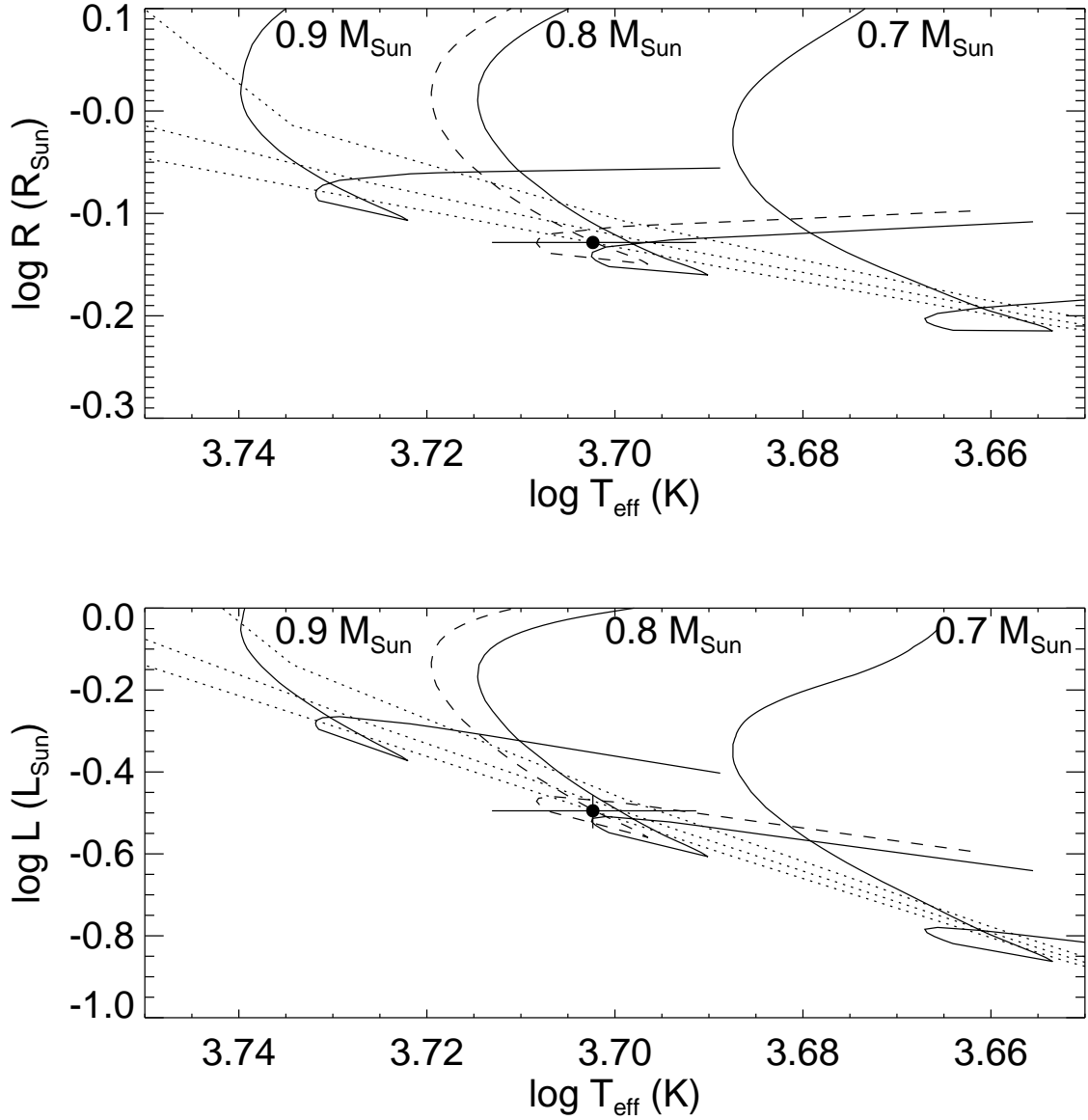}
\caption{H--R diagrams for $\epsilon$ Eri. The solid lines are Y$^2$ isochrones for the stellar masses indicated, the dashed line is for a mass of 0.82 $M_\odot$, and the dotted lines are for a stellar age of 1, 5, and 10 Gyr from the bottom up.}
  \label{yy}
\end{figure}

With our newly calculated $R$, $T_{\rm eff}$, and $L$, we estimated the mass of the central star using the Yonsei--Yale isochrones \citep[Y$^2$,][]{2001ApJS..136..417Y}. Figure \ref{yy} shows the tracks for several masses and the best fit is for a 0.82 $M_\odot$ star. We combined this mass with the mass function ($5.9 \times 10^{-10}\pm 1.0 \times 10^{-10} M_\odot$) and inclination (30.1$\pm$3.8$^\circ$) from \citet{2006AJ....132.2206B} and calculated the planet's mass to be 1.53$\pm$0.22 $M_{\rm Jupiter}$, which is consistent with that listed in Benedict et al. of 1.55$\pm$0.24 $M_{\rm Jupiter}$. The Y$^2$ isochrones also provided an approximate age for the system, which is on the order of 1 Gyr. This is a slightly older age than the results from other studies, which are all less than 1 Gyr \citep[e.g.,][]{1991ApJ...375..722S, 1996AJ....111..439H, 2000ApJ...533L..41S, 2004AandA...426..601D, 2005AandA...443..609S}. We consider our age to be consistent with those measured using other methods because we do not claim high accuracy in that parameter.

We also calculated the size of the system's habitable zone using the following equations from \citet{2006ApJ...649.1010J}:
\begin{equation}
S_{b,i}(T_{\rm eff}) = (4.190 \times 10^{-8} \; T_{\rm eff}^2) - (2.139 \times 10^{-4} \; T_{\rm eff}) + 1.296
\end{equation}
and 
\begin{equation}
S_{b,o}(T_{\rm eff}) = (6.190 \times 10^{-9} \; T_{\rm eff}^2) - (1.319 \times 10^{-5} \; T_{\rm eff}) + 0.2341
\end{equation}
where $S_{b,i}$($T_{\rm eff}$) and $S_{b,o}$($T_{\rm eff}$) are the critical fluxes at the inner and outer boundaries in units of the solar constant. The inner and outer physical boundaries $r_{i,o}$ in AU were then calculated using
\begin{equation}
r_i = \sqrt{ \frac{L/L_\odot}{S_{b,i}(T_{\rm eff})} } \; \; \; \; \; {\rm and} \; \; \; \; \; r_o = \sqrt{ \frac{L/L_\odot}{S_{b,o}(T_{\rm eff})} }.
\end{equation}
We obtained habitable zone boundaries of 0.50 AU and 1.00 AU. $\epsilon$ Eri's planet has a semimajor axis of 3.39$\pm$0.36 AU \citep{2006AJ....132.2206B}, so there is no chance the planet orbits anywhere near the habitable zone. Our values are slightly smaller than those of Jones et al. (0.572 and 1.123 AU).

$\epsilon$ Eri features a dust ring that was imaged at a wavelength of 850 $\mu$m using the Submillimetre Common User Bolometer Array at the James Clerk Maxwell Telescope by \citet{1998ApJ...506L.133G}. They found the ring has a mass between 0.01 and 0.4 $M_{\rm Earth}$, a mass comparable to the estimated mass of comets orbiting in our Solar System. The disk's emission peaks at 100 $\mu$m \citep{2000MNRAS.314..702D} and its flux compared to the central star is on the order of 10$^{\rm -5}$, a contribution that puts it well outside the dynamical range of the NOI, particularly at the optical wavelengths that the NOI uses. At a distance of 55 AU from $\epsilon$ Eri \citep{2003ApJ...598L..51L}, the ring is also larger than the field of view of the NOI. Additionally, \citet{2007AandA...475..243D} found no evidence of dust close to the star that would influence our visibility measurements.

\acknowledgments

The Navy Optical Interferometer is a joint project of the Naval Research Laboratory and the U.S. Naval Observatory, in cooperation with Lowell Observatory, and is funded by the Office of Naval Research and the Oceanographer of the Navy. This research has made use of the SIMBAD database, operated at CDS, Strasbourg, France. This publication makes use of data products from the Two Micron All Sky Survey, which is a joint project of the University of Massachusetts and the Infrared Processing and Analysis Center/California Institute of Technology, funded by the National Aeronautics and Space Administration and the National Science Foundation.

%%%%%%%%%%%%%%%%%%%%%%%%%% Observing Log %%%%%%%%%%%%%%%%%%%%%%%%%%
%
%\begin{deluxetable}{lcl}
%\tablewidth{0pc}
%%\tabletypesize{\scriptsize}
%\tablecaption{Observing Log.\label{obs_log}}
%\tablehead{
% \colhead{$\#$} & \colhead{$\#$}  & \colhead{} \\
% \colhead{Date} & \colhead{Scans} & \colhead{Baseline} }
%\startdata
%2010 Nov 18 & 14 & AC-E6, AC-W7, E6-W7 \\
%2010 Dec 8  &  9 & AC-E6, AC-W7, AE-W7, E6-W7 \\
%2011 Jan 13 & 72 & AC-AE, AC-E6, AC-W7, AE-W7, E6-W7 \\
%\enddata
%\tablecomments{$^\ddagger$The maximum baseline lengths shown in in column 3 are AC-AE 18.9 m, AC-E6 34.3 m, AC-W7 51.5 m, AE-W7 64.4, and E6-W7 79.4 m.}
%\end{deluxetable}
%
%\clearpage

%%%%%%%%%%%%%%%%%%%%%%%%%%% Calibrator Info %%%%%%%%%%%%%%%%%%%%%%%%%%%%

\begin{deluxetable}{lccl}
\tablewidth{0pc}
%\tabletypesize{\scriptsize}
\tablecaption{SED Input Parameters.\label{sed_params}}
\tablehead{ \colhead{Parameter} & \colhead{38 Eri} & \colhead{$\epsilon$ Eri} & \colhead{Source} }
\startdata
$U$ magnitude & 4.53 & 5.19 & \citet{Mermilliod} \\
$B$ magnitude & 4.37 & 4.61 & \citet{Mermilliod} \\
$V$ magnitude & 4.04 & 3.73 & \citet{Mermilliod} \\
$R$ magnitude & 3.84 & 3.19 & \citet{2003AJ....125..984M} \\
$I$ magnitude & 3.68 & 2.74 & \citet{2003AJ....125..984M} \\
$J$ magnitude & 3.43 & 2.23 & \citet{2003tmc..book.....C} \\
$H$ magnitude & 3.25 & 1.88 & \citet{2003tmc..book.....C} \\
$K$ magnitude & 3.21 & 1.78 & \citet{2003tmc..book.....C} \\
$T_{\rm eff}$ (K) & 7079  &  & \citet{all99}  \\
log $g$ (cm s$^{-2}$) & 3.66 &  & \citet{all99} \\
A$_V$ (mag) & 0.05 & & \citet{1980AandAS...42..251N} \\
$T_{\rm eff}$ (K) &  & 5156 & \citet{2004AJ....127.1227C}  \\
log $g$ (cm s$^{-2}$) & & 4.57 & \citet{2004AJ....127.1227C} \\
E($B-V$) & & 0.001 & \citet{2004AJ....127.1227C} \\
$\theta_{\rm LD}$ (mas) & 0.86$\pm$0.02 & 2.04$\pm$0.11 & Calculated here \\
\enddata
\end{deluxetable}

\clearpage

%%%%%%%%%%%%%%%%%%%% Calibrated Visibilities %%%%%%%%%%%%%%%%%%%%

\begin{deluxetable}{cccccl}
\tablewidth{0pc}
\tablecaption{Calibrated Visibilities.\label{calib_visy}}

\tablehead{\colhead{ }    & \colhead{}           & \colhead{Spatial Freq}           & \colhead{ }     & \colhead{ } & \colhead{Baseline} \\
           \colhead{Date} & \colhead{JD-2450000} & \colhead{(10$^6$ cycles/radian)} & \colhead{$V^2$} & \colhead{$\sigma V^2$} & \colhead{Name} \\ }
\startdata

18 Nov 2010 & 5518.803 & 55.286 & 0.399 & 0.013 & AC-W7 \\
 & 5518.803 & 57.136 & 0.380 & 0.013 & AC-W7 \\
 & 5518.803 & 42.157 & 0.679 & 0.029 & AC-E6 \\
 & 5518.803 & 59.105 & 0.408 & 0.012 & AC-W7 \\
 & 5518.803 & 44.924 & 0.643 & 0.026 & AC-E6 \\
 & 5518.803 & 62.985 & 0.338 & 0.015 & AC-W7 \\
 & 5518.803 & 46.249 & 0.616 & 0.038 & AC-E6 \\
 & 5518.803 & 64.842 & 0.341 & 0.016 & AC-W7 \\
 & 5518.803 & 47.660 & 0.607 & 0.031 & AC-E6 \\
 & 5518.803 & 66.822 & 0.268 & 0.016 & AC-W7 \\
 & 5518.803 & 50.308 & 0.510 & 0.039 & AC-E6 \\
 & 5518.803 & 70.533 & 0.274 & 0.018 & AC-W7 \\
 & 5518.803 & 51.603 & 0.525 & 0.045 & AC-E6 \\
 & 5518.803 & 72.348 & 0.241 & 0.021 & AC-W7 \\

\enddata
\tablecomments{A portion of the table is shown here as an example. The rest of the table is available in the online version of \emph{The Astrophysical Journal}. Approximate baseline lengths are AC-E6 34.3 m and AC-W7 51.5 m.}
\end{deluxetable}

\clearpage

%%%%%%%%%%%%%%%%%%%%%%% eps Eri Parameters & Results %%%%%%%%%%%%%%%%%
\begin{deluxetable}{lcl}
\tablewidth{0pc}
%\tabletypesize{\scriptsize}
\tablecaption{$\epsilon$ Eridani Stellar Parameters.\label{parameters}}
\tablehead{ \colhead{Parameter} & \colhead{Value} & \colhead{Reference} }
\startdata
[Fe/H]            & -0.13$\; \pm \;$0.04 & \citet{2004AandA...415.1153S} \\
$V$ magnitude     & 3.73$\; \pm \;$0.01 & \citet{Mermilliod} \\
$K$ magnitude     & 1.78$\; \pm \;$0.29 & \citet{2003tmc..book.....C} \\
$A_{\rm V}$       & 0.00 & \citet{2009ApJ...694.1085V} \\
BC                & 0.20$\; \pm \;$0.10 & \citet{1999AandAS..140..261A} \\
Parallax (mas)	  & 311.73$\; \pm \;$0.11 & \citet{2006AJ....132.2206B} \\
Luminosity ($L_\odot$) & 0.32$\; \pm \;$0.03  & Calculated here \\
$F_{\rm BOL}$ (10$^{-8}$ erg s$^{-1}$ cm$^{-2}$) & 98.0$\; \pm \;$9.3 & Calculated here \\
$\theta_{\rm UD}$ (mas) & 2.080$\; \pm \;$0.028 (1$\%$) & Measured here \\
$\theta_{\rm LD}$ (mas) & 2.153$\; \pm \;$0.028 (1$\%$) & Measured here \\
$R_{\rm linear}$ ($R_\odot$) & 0.74$\; \pm \;$0.01 (1$\%$) & Calculated here \\
$T_{\rm eff}$ (K) & 5039$\; \pm \;$126 (2$\%$) & Calculated here \\
$M$ ($M_\odot$) & 0.82$\; \pm \;$0.05 (6$\%$) & Determined using Y$^2$ model \\
Age & $\sim$1 Gyr & Estimated using Y$^2$ model \\
\enddata
%\tablecomments{}
\end{deluxetable}

\clearpage

%%%%%%%%%%%%%%%%%%%%%%% Figures %%%%%%%%%%%%%%%%%

\clearpage

\clearpage

\clearpage

\clearpage

\end{document}